# An anti-maser for quantum-limited cooling of a microwave cavity

Aharon Blank,[1*] Alexander Sherman[1], Boaz Koren[1] and Oleg Zgadzai[1]

[1] Schulich Faculty of Chemistry, Technion – Israel Institute of Technology, Haifa, 3200003, Israel

*Corresponding author. Email: ab359@technion.ac.il

**Abstract:** The maser, a microwave (MW) analog of the laser, is a well-established method for generating and amplifying coherent MW irradiation with ultra-low noise. This is accomplished by creating a state of population inversion between two energy levels separated by MW frequency. Thermodynamically, such a state corresponds to a small but negative temperature. The reverse condition, where only the lower energy level is highly populated, corresponds to a very low positive temperature.  In this work, we experimentally demonstrate how to generate such a state in condensed matter at moderate cryogenic temperatures.  This state is then used to efficiently remove microwave photons from a cavity, continuously cooling it to the quantum limit, well below its ambient temperature.  Such an "anti-maser" device could be extremely beneficial for applications that would normally require cooling to millikelvin temperatures to eliminate any MW photons. For instance, superconducting MW quantum circuits (such as qubits and amplifiers) could, with the use of this device, operate efficiently at liquid helium temperatures.

**One-Sentence Summary:** Cooling the microwave mode of a cavity by coupling it to an optically-pumped spin system that maintains an ultra-low, quantum-limited spin temperature.

Cooling is vital in numerous scientific and industrial applications and utilizes various mechanisms. For example, mechanical refrigeration, a common method, operates on the Carnot cycle, transferring heat from one area to another. For ultra-low temperatures, liquid helium is used, which cools to 4.2 Kelvin and even lower when made into a superfluid. Alternatively, laser cooling employs the Doppler shift to slow and cool atoms with precision-tuned laser beams, enabling ultra-cold experiments like Bose-Einstein condensates and atomic clocks (*1*). Magnetic cooling, another process, uses the magnetocaloric effect, causing a material's temperature to change with an altered magnetic field, and is mostly applied in low-temperature refrigeration (*2*). Lastly, thermoacoustic cooling uses sound waves in a pressurized gas to shift heat, generating a cooling effect without refrigerants (*3*).

One recent field that makes use of extreme cooling capability is superconducting circuits operating within the microwave (MW) domain, which have emerged as frontrunners in quantum science and technology applications. Their uses span a broad range, from superconducting quantum bits (qubits) to single MW photon detectors, quantum sensors, and devices for quantum communication (*4, 5*). A common requirement across all these applications is the necessity to cool these devices to temperatures of approximately 100 mK or less, to prevent interaction with surrounding MW photons. For instance, at 10 GHz, $h\nu \sim k_B T$ when $T \approx 0.47$ K ($h$ is Planck's constant, $\nu$ is the frequency, $k_B$ is Boltzmann's constant, and $T$ is the temperature). Hence, to avoid thermal excitation of a two-level system operating in this frequency regime, it needs to be cooled well below that temperature so the qubit is almost entirely in the ground state.

Cryocoolers capable of reaching mK temperatures, based on dilution refrigerators, have been in experimental use for over two decades (*6*). However, they pose significant financial and operational challenges due to their high cost, reliance on the rare $^3$He isotope, and limited cooling capacity. These factors constrain the practical dissemination and scalability of quantum devices based on superconducting circuits operating in the MW regime.

Superconductivity itself is not restricted to mK temperatures, and superconductors with desirable quantum properties, such as Nb, or NbTiN, function well even at around 10 K (*7, 8*). It therefore would be highly desirable to operate superconducting MW quantum devices at these higher temperatures without concern for the MW photons that could potentially disrupt their operation.

In this paper, we demonstrate the use of spin-based cooling approach as a method that could potentially be used achieve this ideal scenario. We employ a 3-level solid-state quantum system, specifically, the negatively charged nitrogen-vacancy (NV⁻) defect in a diamond single crystal (Fig. 1). This system exhibits the remarkable ability to use optical irradiation to selectively populate the central ($|0\rangle$) level while depleting the lower ($|-1\rangle$) and upper ($|1\rangle$) levels. Previous studies have leveraged this system to generate coherent MW irradiation and demonstrate low-noise amplification of weak MW signals (*9, 10*), by tuning its operation to the $|0\rangle \leftrightarrow |-1\rangle$ transition, which showcases population inversion—i.e., maser action. In contrast, our current work utilizes the $|0\rangle \leftrightarrow |1\rangle$ transition of the NV⁻ system, in which only the lower level is populated. This results in the removal of photons from the cavity that houses the system. Namely, such "anti-maser" action suppresses, rather than generates or amplifies, MW irradiation, effectively cooling the microwave mode of the cavity well below its ambient operating temperatures — into the quantum regime with less than one MW photon in this frequency. Pairing this anti-maser cooling device with the abundance of microwave superconducting quantum devices available today (*5*) could enable their useful operation even at temperatures of ~4.2-10 K. This operation range is significantly easier to achieve, maintain, and compatible with other heat-generating devices, such as laser light sources and high electron mobility transistors (HEMTs) (*11*), used in a variety of quantum-microwave related experiments.

It's worth noting that recent efforts have demonstrated some level of microwave (MW) mode cooling capabilities using an approach similar to the one outlined in this paper, though with considerable limitations compared to what we've now accomplished. For instance, in a previous study using optically-pumped pentacene molecules at room temperature, the MW mode of a dielectric resonator was cooled to ~ 50 K at 1.45 GHz, but the effect lasted for less than 1 ms (*12*). Another study involving NV⁻ centers in diamond demonstrated that at room temperature and 2.87 GHz, the MW mode could be cooled to ~188 K for 10 ms pulses (*13*). One paper theorized that a similar system could theoretically cool the MW mode to ~116 K for durations of 20 ms in pulsed mode at 9.2 GHz and room temperature (*14*). Furthermore, a recent arxiv paper, which has not yet undergone peer review, presented experimental results demonstrating steady-state cooling to ~ 150 K for an NV⁻ center system at room temperature and ~ 3 GHz (*15*). While these achievements represent significant progress in this field and underscore the importance of developing such cooling methodologies, the limited duration of the cooling process and the rather high MW mode noise temperatures achieved do not offer a substantial advantage over traditional cooling schemes. We contend that the largest technological and experimental gap in cooling capacity exists between ~ 4.2 K and ~ 0.5 K and below. This means that MW mode cooling methodologies need to bridge this gap and reach a regime where there is a single noise photon or fewer in the cavity. This would significantly impact the field of superconducting MW circuits and similar quantum-regime MW experiments. In this work, we demonstrate a configuration that can cool the MW mode to this single photon noise regime when operating at a temperature of 5K. This makes it much more viable for practical quantum MW applications.

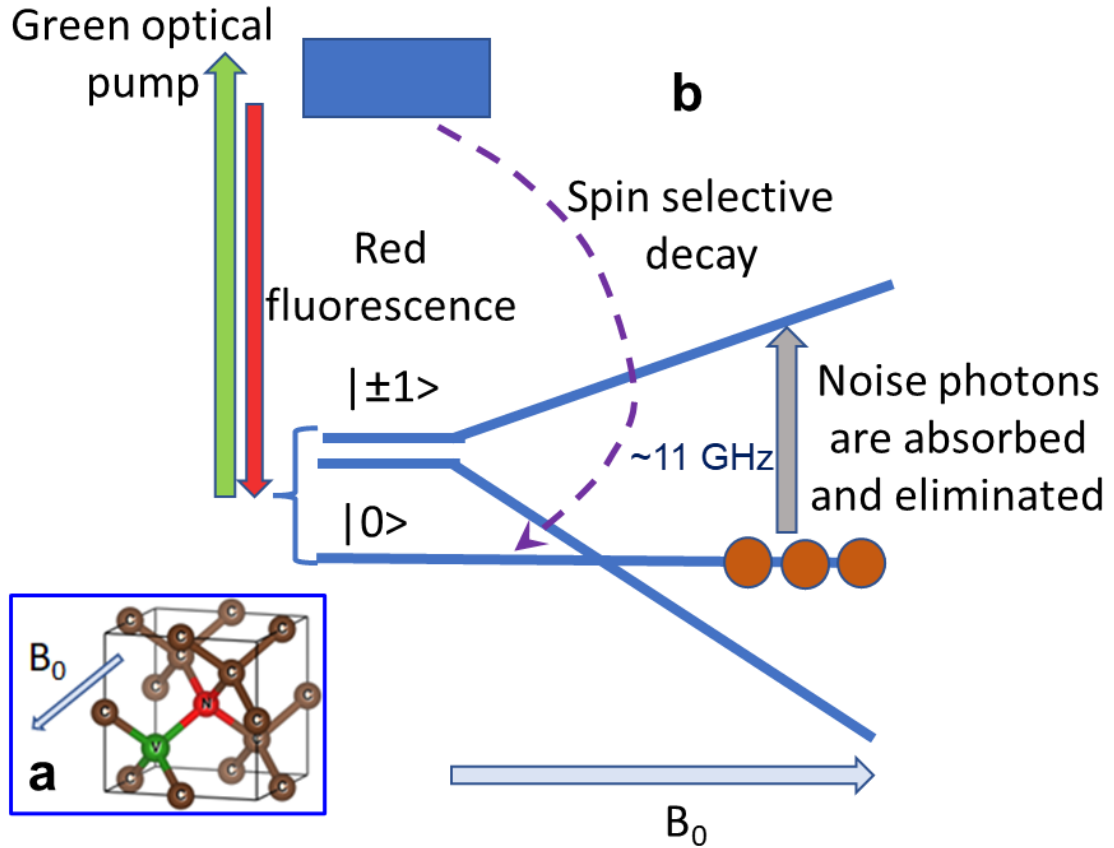

**Figure 1:** The NV⁻ centre in a diamond single crystal. (a) The image depicts the diamond crystal lattice with carbons represented by brown spheres. One of the carbon atoms is replaced by a nitrogen (marked in red) and another nearby carbon is substituted with a vacancy (shown in green). A static magnetic field is applied along the NV⁻ axis (denoted by $B_0$). (b) The diagram illustrates the three lowest levels of the NV⁻ system, which are Zeeman-split, and the optically excited states. At zero magnetic field, the $m_s = 0$ level (noted as |0>) has lower energy than the two degenerate $m_s = \pm 1$ levels (|±1>). The application of green light leads to red fluorescence along with other non-radiative spin-selective transitions, resulting in an enhanced population of the |0> state. This selectivity is maintained in the presence of a static magnetic field aligned along the NV⁻ axis. Any existing noise photons in the cavity are then efficiently absorbed by the |0> state, promoting it to the |+1> state. This state is then efficiently depleted through optical pumping and non-radiatively transitions back to the |0> state.

**The anti-maser microwave cooling device**

The schematic of our microwave (MW) mode cooling device is presented in Figure 2a-c, with complete mechanical and electromagnetic details of the cavity provided in the Supplementary Material (SM). Our experiments were conducted at temperatures ranging from 5 to 30 K. At first look, the necessary experiments to monitor the noise in the cavity should be straightforward. One first needs to adjust the static magnetic field to match the frequency of the $|0\rangle \leftrightarrow |1\rangle$ transition at the resonance frequency of the cavity. Following this, the noise spectrum emitted from the cavity should be directly recorded by a spectrum analyzer (SA), preceded by a low-noise amplification chain. Measurements should be taken both with and without light excitation, enabling a comparison of the cavity noise variations under these conditions. Unfortunately, this simplified approach does not perform optimally under our experimental conditions due to the strong collective coupling of the spins to the cavity. This coupling significantly impacts the cavity's MW properties under light irradiation, as described below. Consequently, we utilize a slightly more complex setup, depicted in Figure 2d.

Figure 3 displays the results of noise emission measurements from the cavity across a variety of temperatures and frequencies near the cavity's resonance frequency (of ~10.98 GHz), when a 20 dB directional coupler is used in the setup (the results for 10 dB coupler are given in Fig. S5). The data demonstrate a considerable decrease in emitted noise upon activation of the LEDs, across all tested temperatures. Yet, to correlate this noise reduction to an actual decrease in MW photon noise within the cavity, one must carefully consider the specifics of the measurement setup. An analysis of the noise origins in our setup (Figure 2d) reveals that the noise entering the low noise amplifier (LNA) is a weighted sum of the noise coming from the cavity and noise received through the directional coupler. This noise is then amplified and registered by the SA. However, it's clear that the noise is not uniform (white), and it increases at and around the resonance frequency of the cavity even when light is absent. This can be attributed to three frequency-dependent effects: (i) The noise emitted from the cavity is frequency-dependent, as the coupling of the cavity to the external environment varies with frequency. (ii) The gain of the LNA is affected by the input impedance it encounters, which is frequency-dependent due to the frequency-dependent cavity matching condition. (iii) The noise temperature of the LNA also relies on its input impedance, which is frequency-dependent. Due to the strong coupling of the spins to the cavity, light also alters the coupling of the cavity to the LNA and potentially affects all three terms mentioned above, in addition to reducing the internal MW noise of the cavity. Therefore, understanding these effects and their possible influence on our measurements is critical to accurately quantify the level of noise in the cavity under light irradiation. Without addressing these issues, one might incorrectly infer, for instance, that the observed noise reduction in Figure 3 when light is applied could simply be due to a decrease in LNA gain as the impedance of the cavity changes or changes in the cavity coupling the output port, disregarding the actual reduction in cavity noise. In the following section and in the SM we provide more details as to how we actually quantify the noise temperature of the MW mode inside the cavity under light illumination.

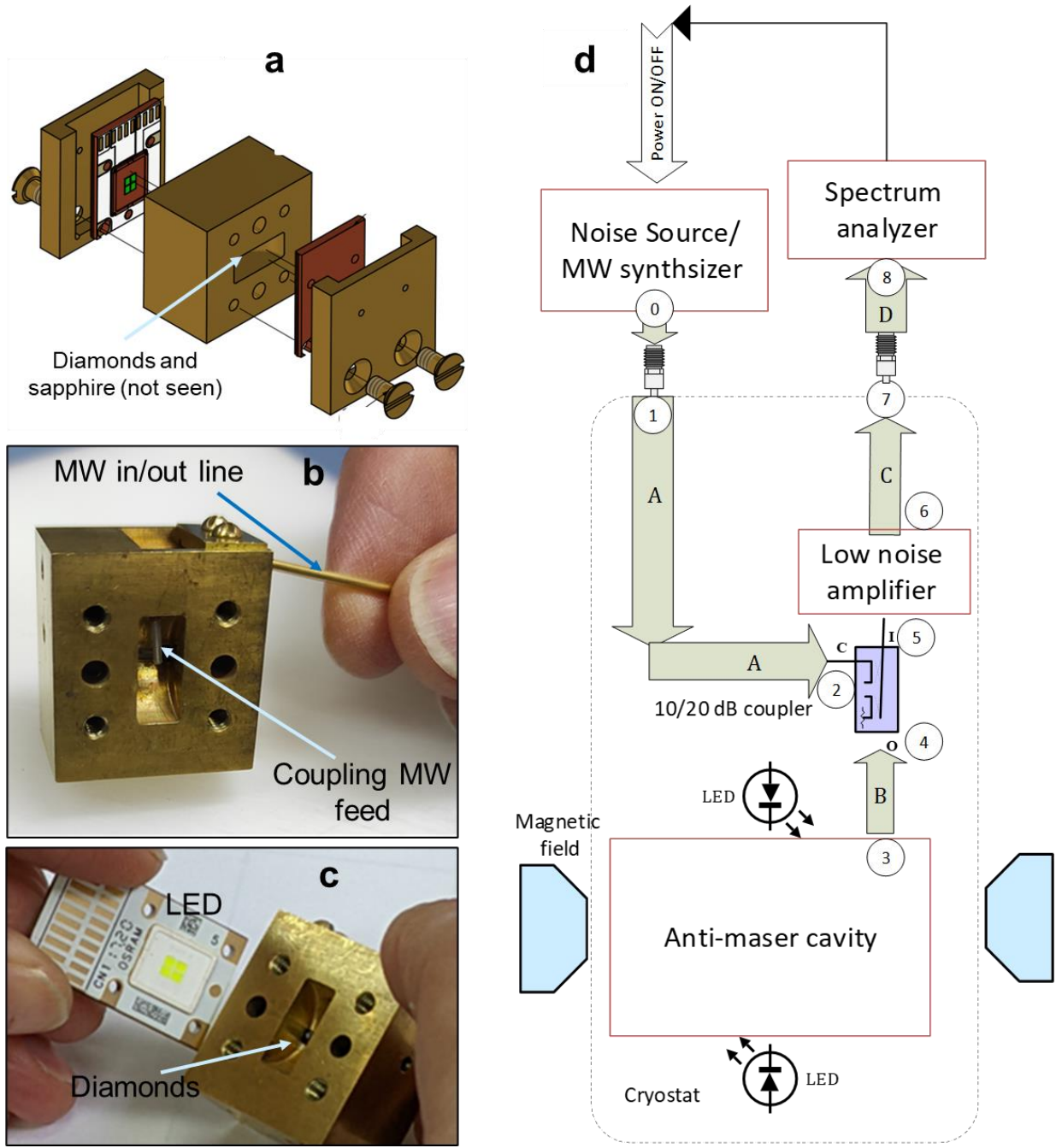

**Figure 2:** The anti-maser device and the experimental setup for evaluating its MW noise. (a) A 3D breakdown of the cavity metallic structure, with the LEDs attached next to it. (b) A photo of the cavity of the anti-maser with the MW feed coupling MW energy in/out of it. (c) A photo of the open cavity with one of the LEDs next to it and the diamonds inside it. (d) The microwave setup for measuring the gain of the low noise amplifier (LNA) connected to the cavity output, and measuring the noise emitted from the anti-maser cavity (using the Y-factor method). For LNA gain measurements, point 0 is occupied by a MW synthesizer, while for cavity noise measurements, it is replaced by a calibrated noise source. The numbers on the plot correspond to points along the path of the MW noise/signal, as analysed in the paper (see also Table S1).

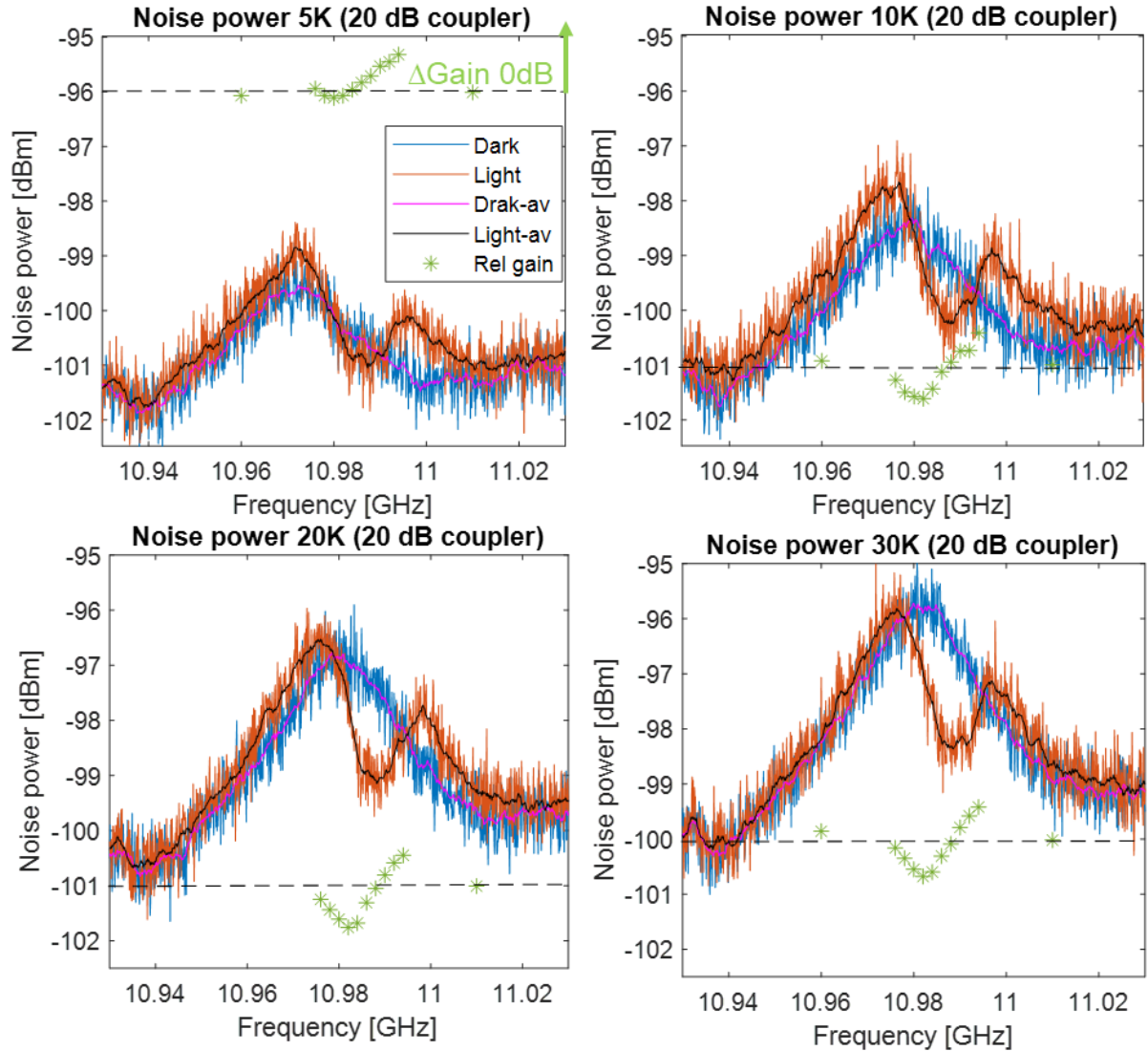

**Figure 3:** Measurement of noise emitted from the anti-maser cavity with a small constant addition of noise coming from the 20 dB coupler of our setup (Fig. 2d – noise source is OFF). The data is shown for the temperature range of 5-30 K under dark and light conditions, when the field is set to match the |0>↔|1> transition to that of the resonance frequency of the cavity. The raw data noise measurements are superimposed with a moving window average. The overall trend for increase noise when the light is ON, especially visible at lower temperature, is due to slight heating (by ~1-2 K max) of the internal parts of the cavity by light. This is not recorded by our nearby temperature sensor that still presents the required temperature for each regime. The green asterisks show the relative change in the gain of the LNA when the light is turned. The scale for the relative gain is in dB according to the left axis but with zero shifted to the dashed line. Legend is the same in all plots.

**Quantifying the of anti-maser cooling effect**

To address the aforementioned concerns and offer a quantitative evaluation of cavity noise under light illumination, we adopt the following approach: Before conducting our noise experiments, we independently measure the cavity's coupling to the external feed line both with and without light illumination, ensuring the magnetic field aligns with the resonance of the $|0\rangle \leftrightarrow |1\rangle$ transition. These findings (refer to Fig. S3) distinctly illustrate the impact of light on the cavity's coupling coefficient, which is effectively reduced. The reason for that is that the anti-maser effect essentially absorbs most MW photons in the resonator, making it effectively more lossy and thus becomes undercoupled. Notably, similar measurements conducted when the magnetic field is off-resonance demonstrated no measurable impact of light on the cavity's coupling coefficient. Our next step involves measuring the change in the gain of the LNA as light is applied to the cavity. We accomplish this by injecting a microwave signal through the directional coupler and measuring the results after the LNA amplification, using the SA. This is done by placing a MW synthesizer source in point 0 shown in Fig. 2d. The observed variations in gain, approximately 0.5 dB, are marked by the green asterisks in Fig. 3. Lastly, we estimate the impact of changes in input impedance on the noise temperature of the LNA. Measurements of this effect cannot be carried out in our setup and therefore we resort to calculated data (Fig. S6). These indicate that a non-ideal input impedance can increase the noise temperature of the LNA by up to a factor of approximately 2 within the relevant regime where the cavity coupling coefficient changes. However, since this noise is also contingent on the phase of the input impedance - which is challenging to quantify accurately in our setup - we omit from our analysis the potential increase in LNA noise temperature when light is applied (due to coupling changes). This implies that our results would ***overestimate*** the cavity's noise, and in reality, the noise can only be equal or ***lower*** than our measurements and corresponding data analysis suggest.

We use the collected data regarding all these parameters to estimate the noise temperature of the cavity. This noise temperature measurement is executed using the Y-factor method (*16*) with the configuration depicted in Fig. 2d. (When the noise source is applied in point 0, refer to Supplementary Material for details.) The measurements were conducted using both a 10 dB directional coupler and a 20 dB coupler, each of which offers unique advantages and disadvantages when measuring LNA gain variations and the Y-factor. The outcomes are presented in Fig. 4. We take advantage of the fact that the noise temperature of the cavity when light is OFF is known (simply the temperature of the cavity), and all we seek is the *change* in this value when light is turned ON. This enables us to enhance the precision of our measurements by relating to the *variation* in noise emitted from the cavity when light is applied by focusing on the *change* in the Y-factor for the light OFF/ON conditions (see also SM). The findings from these measurements, after taking into account all relevant analyses, are furnished in Table 1. (Also, refer to Table S1 in the Supplementary Material.) The results unambiguously demonstrate a significant decrease in cavity noise, often by a factor of 2 or more in comparison to the cavity's ambient temperature. The measured data is juxtaposed with theoretical results, which are grounded in a numerical simulation of NV$^-$ energy levels within the cavity. The simulations are based on a numerical model we have recently developed for maser device analysis (*10*), with minor changes, so that the results relate to the |0⟩↔|1⟩ anti-maser transition and not to the |-1⟩↔|0⟩ maser transition. The simulation is written in MATLAB$^{TM}$ and is available online in Zenodo or from the corresponding author. The simulated results also enable us to calculate the internal cavity noise temperature in the case of dismissingly small coupling to the outside port (Table 1), which is ~0.63 K.

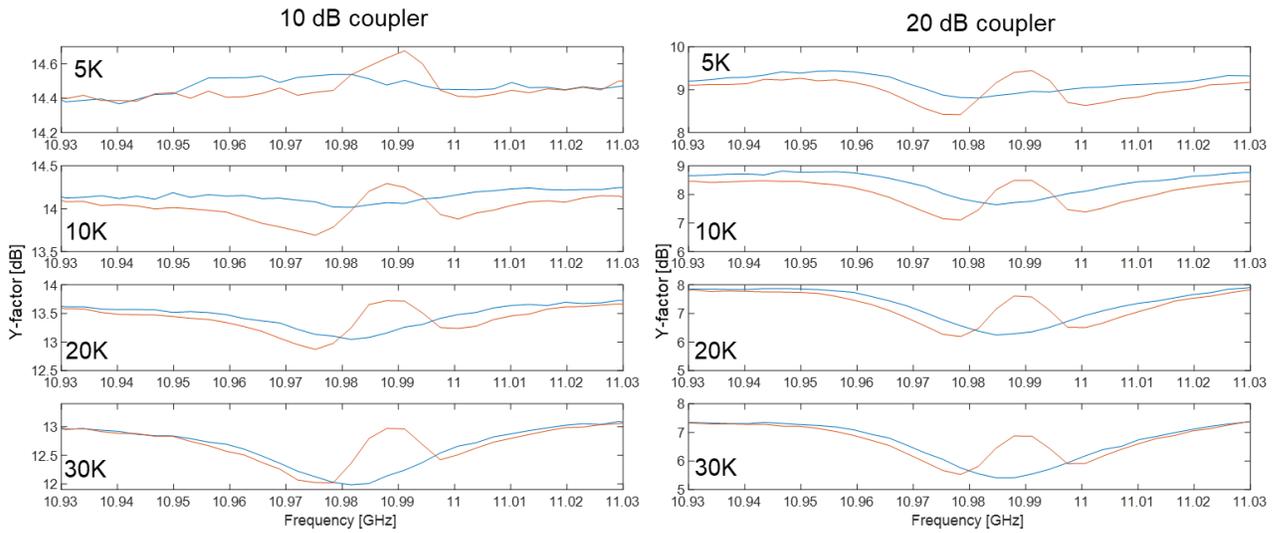

**Figure 4:** Measurement of the Y-factor (in dB) using the setup of Fig. 2d for the temperatures in the range of 5-30 K and with 10 and 20 dB directional couplers. The blue lines correspond to the condition when the LEDs are OFF and the red when LEDs are ON.

| Parameter | | ΔNoise @SA (dB) | ΔLNA Gain (dB) | ΔY (dB) | S$_{11}$ (dB) light OFF; light ON and on resonance | Cavity noise (K), meas. | Cavity noise (K), calc. (w/o coupling) |
|---|---|---|---|---|---|---|---|
| **10dB** | 5K | -0.51±0.1 | -0.31±0.04 | 0.17±0.01 | -11.9±0.2; -3.8±0.2 | ***2.5±0.14*** | ***1.07 (0.63)*** |
| coupler | 10K | -0.87±0.1 | -0.31±0.04 | 0.22±0.01 | -14.8±0.2; -4.7±0.2 | 6.5±0.14 | 1.92 (1.09) |
| | 20K | -1.13±0.1 | -0.53±0.04 | 0.57±0.01 | -16.2±0.2; -6.6±0.2 | 10.5±0.15 | 3.42 (2.0) |
| | 30K | -1.22±0.1 | -0.34±0.04 | 0.83±0.01 | -20±0.2; -7.2±0.2 | 13.4±0.15 | 4.72 (2.93) |
| 20dB coupler | 5K | -0.29±0.1 | +0.28±0.04 | 0.51±0.01 | -11.9±0.2; -3.8±0.2 | ***3±0.1*** | ***1.07 (0.63)*** |
| | 10K | -1.08±0.1 | +0.06±0.04 | 0.78±0.01 | -14.8±0.2; -4.7±0.2 | 5.6±0.1 | 1.92 (1.09) |
| | 20K | -1.68±0.1 | -0.05±0.04 | 1.32±0.01 | -16.2±0.2; -6.6±0.2 | 8±0.1 | 3.42 (2.0) |
| | 30K | -1.99±0.1 | -0.09±0.04 | 1.46±0.01 | -20±0.2; -7.2±0.2 | 10.3±0.1 | 4.72 (2.93) |

**Table 1:** Summary of the measured and calculated results for the characteristics of the anti-maser device. Data is presented for 10 and 20 dB couplers at temperatures of 5, 10, 20, and 30K. The first category (ΔNoise @SA) indicates the noise reduction measured by the SA when the LEDs are activated, and the static field is on resonance (based on Figs. 3 and S5). The second category (ΔLNA Gain) represents the change in LNA gain upon the activation of the light when the static field is on resonance (based on the data shown Figs. 3 and S5 by green asterisks). The third (ΔY) is the Y-factor alteration upon light activation under resonance condition (based on Fig. 4). The fourth category presents the $S_{11}$ of the cavity under the light OFF/ON conditions conditions (Fig. S3). The fifth reveals the measured cavity noise, deduced from the ΔY analysis, as detailed in the main text. The sixth and final category illustrates the calculated cavity noise temperature, derived by our theoretical model. The numbers in parentheses denote the calculated noise temperature when the cavity's coupling to the external port is infinitesimally small. The most important t results of this paper are emphasized by a bold italic font.

**Discussion**

We have realized a unique type of spin refrigerator. Selective cooling of a ~11 GHz MW cavity mode has been achieved with an average number of photons being less than 5 in our 5 K experiments ($k_BT/h\nu$ ~4.7, for the 2.5 K noise temperature we measured). This number is an upper limit since our analysis did not consider the possible increase in LNA noise temperature due to the non-optimal impedance it "sees" at its input when the NV spins are pumped. As can be seen in Table 1, these results agree with our theoretical model, which predicted in all the temperatures we tested a slightly lower noise than our experimental upper limit. Much of the noise of our cavity comes due to its coupling to its in/out MW port. Such coupling is unavoidable if one wants to measure the cavity's noise parameters. However, in principle the cavity can be very weakly coupled to the outside and still be useful for housing quantum MW circuits in it. Our theoretical model enables us to predict the expected noise temperature of the cavity in such a s case of minimal coupling to the outside world, reaching ~0.63 K for 5 K ambient temperature, *meaning potentially having just slightly over 1 photon of noise in average in the cavity mode*. These noise levels can be achieved for a rather low $Q$ cavity ($Q$ ~320) and with a large bandwidth (~10 MHz) of operation, relying on the relatively broad spectral lines of the NVs. This gives much flexibility in combining the anti-maser cooling device with superconducting quantum circuits. In principle, for a bandwidth of 10 MHz, one can envision the more optimal use of a higher $Q$ cavity, leading to even reduced noise level of the MW mode (for example, for a $Q$ of 1000, the calculated achievable cavity noise temperature is ~ 0.2 K, when coupling to the outside is minimal). It should be noted that the spin temperature by itself in our experiments was approaching 0 K (fully populated $|0>$ state) for all temperatures below 30 K (see Fig. S4).

The capability to operate at ~ 5 K temperature while eliminating almost all noise photons from the cavity MW mode is first and foremost important to applications involving superconducting quantum MW circuits. For such devices, one may envision the use of a closed cavity as we describe here, with the superconducting circuits placed directly in it. As noted above, such architecture has the advantage of minimizing noise contribution from the outside and minimize coupling losses. These and additional disturbing effects exist in other possible mode cooling alternatives, for example where a cold load is coupled to a warm cavity (*17, 18*) or when using Rydberg atoms to remove thermal photons from millimetre-wave cavities via stimulated absorption (*19*). Our spin refrigerator device can also be useful to eliminate photons that still exist even in ~10 mK ambient temperature due to external connections of superconducting MW devices, which decrease the coherence time of the device. Existing solution to this problem (e.g. in (*20*)) come on the expense of reducing detection sensitivity.

In addition to enableing the operation of MW superconducting quantum circuits at elevated temperatures, our methodology can be of importance to applications where heat dissipation in mK coolers is a major issue, such as electro-optical quantum state transduction (*21*) and microwave-optical photon entanglement generation based on hybrid superconducting systems (*22*), which are all important steps towards the realization of scalable quantum networks (*23*). In such applications, a high-power optical pump is typically required to boost the transduction efficiency or photon entanglement generation rate, while heating due to optical absorption by the dielectrics that host the optical modes becomes significant.

**Acknowledgments:**

We thank IAI ELTA Systems' Quantum Program for its support in this research and Jörgen Stenarson from Low Noise Factory for supplying us with the noise temperature simulation of the LNA.

**Funding:**

This project was funded by the Israel Science Foundation (ISF), Grant No. 1357/21, and the Israel Innovation Authority, Grant No. 67697 with ELTA Systems. This research was partially supported by the Technion's Hellen Diller Quantum Center and its Russell Berrie Nanotechnology Institute.

**Authors' contributions:**

All authors discussed the results and approved the manuscript. A.B. conceived the project. A.B. and A.S. carried out the methodological experimental work. A.B. and O.Z. and B.K prepared the figures visualization. Funding was acquired by A.B., who also supervised the research. The original draft was written by A.B and the final review and editing was carried out by A.B, and O.Z.

**Competing interests:** The authors have no competing interests.

**Data and materials availability:** All data needed to evaluate the conclusions in the paper is available in the main text or the Supplementary Materials section. Additional software used to simulate and analyze the results is deposited in Zenodo.

**References**

1. W. D. Phillips, Nobel Lecture: Laser cooling and trapping of neutral atoms. *Rev Mod Phys* **70**, 721-741 (1998).
2. V. Franco *et al.*, Magnetocaloric effect: From materials research to refrigeration devices. *Progress in Materials Science* **93**, 112-232 (2018).
3. N. A. Zolpakar, N. Mohd-Ghazali, M. Hassan El-Fawal, Performance analysis of the standing wave thermoacoustic refrigerator: A review. *Renewable and Sustainable Energy Reviews* **54**, 626-634 (2016).
4. X. Gu, A. F. Kockum, A. Miranowicz, Y.-x. Liu, F. Nori, Microwave photonics with superconducting quantum circuits. *Phys. Rep.* **718-719**, 1-102 (2017).
5. M. Casariego *et al.*, Propagating quantum microwaves: towards applications in communication and sensing. *Quantum Science and Technology* **8**, 023001 (2023).
6. J. Ekin, *Experimental Techniques for Low-Temperature Measurements: Cryostat Design, Material Properties and Superconductor Critical-Current Testing*. (Oxford University Press, 2006).
7. R. Hott, R. Kleiner, T. Wolf, G. Zwicknagl, in *digital Encyclopedia of Applied Physics*. pp. 1-55.
8. A. Anferov, K.-H. Lee, F. Zhao, J. Simon, D. I. Schuster, Improved Coherence in Optically-Defined Niobium Trilayer Junction Qubits. *arXiv:2306.05883 [quant-ph]*, (2023).
9. J. D. Breeze, E. Salvadori, J. Sathian, N. M. Alford, C. W. M. Kay, Continuous-wave room-temperature diamond maser. *Nature* **555**, 493-+ (2018).
10. A. Sherman *et al.*, Diamond-based microwave quantum amplifier. *Science Advances* **8**, eade6527 (2022).
11. M. W. Pospieszalski, in *2018 22nd International Microwave and Radar Conference (MIKON)*. (2018), pp. 1-6.
12. H. Wu, S. Mirkhanov, W. Ng, M. Oxborrow, Bench-Top Cooling of a Microwave Mode Using an Optically Pumped Spin Refrigerator. *Phys. Rev. Lett.* **127**, 053604 (2021).
13. W. Ng, H. Wu, M. Oxborrow, Quasi-continuous cooling of a microwave mode on a benchtop using hyperpolarized NV− diamond. *Appl. Phys. Lett.* **119**, (2021).
14. Y. Zhang *et al.*, Microwave mode cooling and cavity quantum electrodynamics effects at room temperature with optically cooled nitrogen-vacancy center spins. *Npj Quantum Inform* **8**, 125 (2022).
15. D. P. Fahey *et al.*, Steady-state microwave mode cooling with a diamond NV ensemble. *arXiv:2203.03462 [quant-ph]*, (2022).
16. Agilent. (2014).
17. B. Albanese *et al.*, Radiative cooling of a spin ensemble. *Nature Physics* **16**, 751-755 (2020).
18. M. Xu *et al.*, Radiative Cooling of a Superconducting Resonator. *Phys. Rev. Lett.* **124**, 033602 (2020).
19. S. Haroche, J. M. Raimond, in *Advances in Atomic and Molecular Physics,* D. Bates, B. Bederson, Eds. (Academic Press, 1985), vol. 20, pp. 347-411.

20. Z. Wang *et al.*, Cavity Attenuators for Superconducting Qubits. *Physical Review Applied* **11**, 014031 (2019).
21. L. Fan *et al.*, Superconducting cavity electro-optics: A platform for coherent photon conversion between superconducting and photonic circuits. *Science Advances* **4**, eaar4994 (2018).
22. A. Rueda, W. Hease, S. Barzanjeh, J. M. Fink, Electro-optic entanglement source for microwave to telecom quantum state transfer. *Npj Quantum Inform* **5**, 108 (2019).
23. H. J. Kimble, The quantum internet. *Nature* **453**, 1023-1030 (2008).

Supplementary Materials for

**An anti-maser for quantum-limited cooling of a microwave cavity**

Aharon Blank,[1*] Alexander Sherman[1], Boaz Koren[1] and Oleg Zgadzai[1]

[1] Schulich Faculty of Chemistry, Technion – Israel Institute of Technology, Haifa, 3200003, Israel

*Corresponding author. Email: ab359@technion.ac.il

**The PDF file includes:**

**Texts S1 to S4**

**Figures S1 to S6**

**Table S1**

**References to the SM**

## S1. Materials and methods

*Diamond single crystals*: The diamonds, procured from Chenguang Machinery & Electric Equipment Co., Ltd., China, were manufactured using the high-pressure-high-temperature (HPHT) technique. They are shaped as rectangle prims, having dimensions of $3 \times 2 \times 1$ mm and a [1 1 1] crystallographic face orientation. Their nitrogen impurity concentration is approximately 100–200 ppm, with inhomogeneity observed throughout the diamonds. The concentration of substitutional nitrogen (paramagnetic P1 centers) is estimated at ~50–100 ppm, as measured by continuous wave ESR using Bruker's X-band EMX system. Initially, the diamonds had a negligible native concentration of NVs (<0.1 ppm), necessitating electron irradiation (5 MeV) at a dose of ~$1 \times 10^{19}$ electrons/cm$^2$. This process was performed at Sorvan Ltd., Yavneh, Israel, to convert P1s to NVs. The irradiation was carried out in two phases: the initial 135 hours without heating, followed by an additional 70 hours with heating up to ~750 °C. The diamonds were stored in a vacuum within sealed quartz tubes during this procedure. Following irradiation, the diamonds were cleaned using a boiling solution of nitric (68%), sulfuric (98%), and perchloric (70%) acids mixed in a 1:1:1 ratio for one hour. This cleaning step is vital for eliminating any carbonization residues, which reduces dielectric losses and subsequently enhances the cavity's $Q$ levels. The diamond pieces used in the device exhibited an NV$^-$ concentration of approximately 20 ppm. This data was obtained by performing pulsed ESR Hahn echo field sweeps using our pulsed ESR spectrometer and comparing the NV signal to that of the P1s. The same field-swept echo data allowed us to ascertain the NVs' $T_2^*$, which measured approximately 35 ns (equivalent to a linewidth of ~ 9 MHz).

*Anti-maser cavity:* Figure S1a displays the cavity housing the diamonds. This cavity is specifically engineered to deliver a high filling factor for the diamonds, meaning that the microwave (MW) magnetic field is primarily concentrated within the diamonds. It's also designed to enable efficient optical excitation and minimize dielectric losses, meaning the MW electric field is largely external to the diamonds and concentrated in the sapphire. These properties are illustrated in Fig. S1f-g, which portrays the computed E and H fields at the center of the cavity. The overall structure of the cavity bears resemblance to the one discussed in (*1*), albeit with adjusted dimensions and enhancements in the optical excitation methodology. The cavity, fabricated from gold-plated brass, assumes the shape of two half-cylinders joined by narrow slits.

Two single-crystal diamonds and two sapphire prisms are accommodated within these slits. The cavity's resonance frequency is approximately 11 GHz, and an unloaded $Q$ of ~320 was measured using a vector network analyzer (VNA).

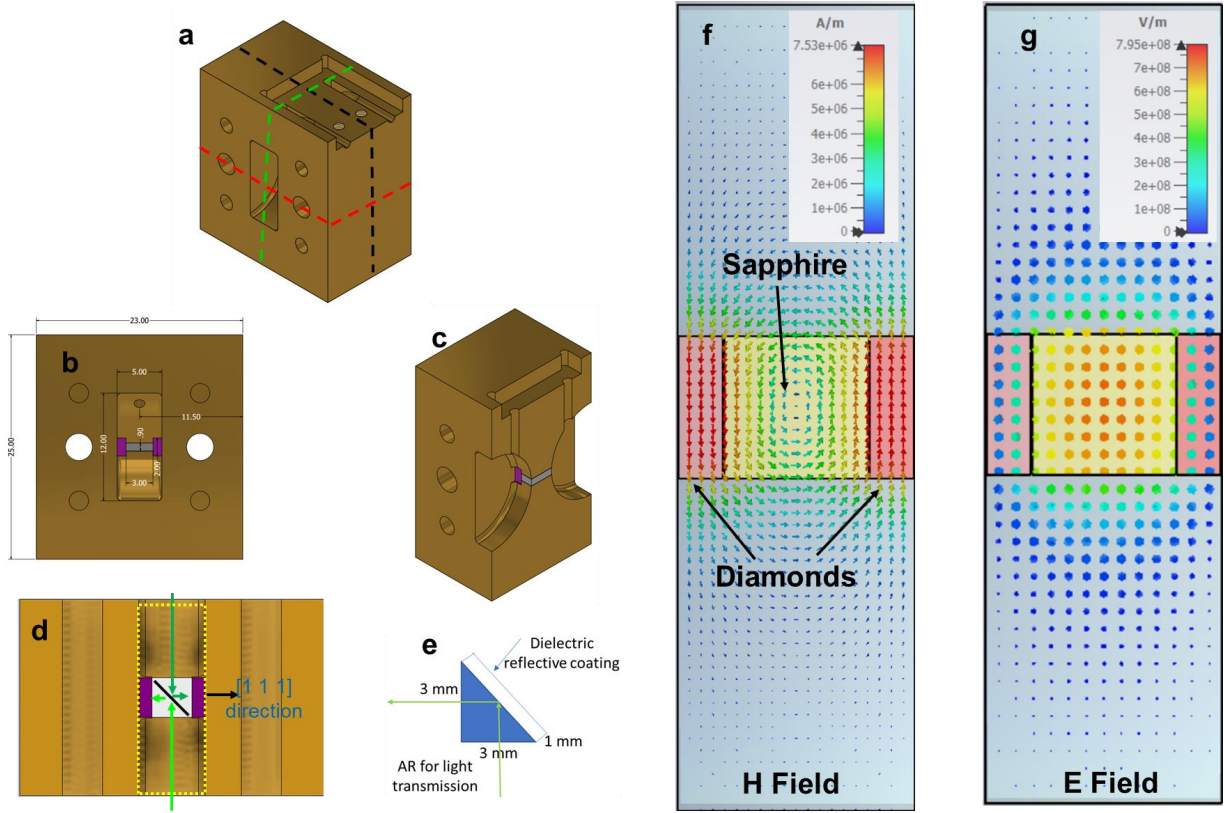

**Figure S1:** Details of the anti-maser cavity structure. (a) An isometric 3D sketch of the main body of the cavity, constructed from gold-plated brass. (b) The dimensions (in mm) of the cavity, diamonds, and sapphire crystals, derived from a cross-section taken along the dashed black line in image (a). (c) An isometric cutaway view of the cavity, diamonds, and sapphire, taken along the dashed green line in image (a). (d) This sketch illustrates the internal structure of the cavity from a cut made along the red dashed line in image (a). It showcases two back-to-back triangle prisms, crafted from sapphire (custom design and fabrication by Glas Opticslens Co., Ltd., China). These prisms are used to efficiently direct green light onto the diamonds' surface. (e) The specifics and dimensions of one of the sapphire prisms utilized in this work. (f) The computed microwave magnetic field in the plane indicated by the yellow dashed line in image (d). (g) The computed microwave electric field in the plane denoted by the yellow dashed line in image (d).

**S2.** *Measurements of the cavity coupling coefficient and the NV⁻ Zeeman levels population:*
The coupling coefficient of the cavity (specifically, its reflection coefficient, $S_{11}$) was evaluated using the setup illustrated in Fig. S2. The cavity was connected directly to a pulsed ESR spectrometer (SpinUp-X model by Spinflex - Israel). This spectrometer has the ability to measure the reflection coefficient as a function of frequency (referred to as "Tune mode") while transmitting very weak pulses with a power of about -85 dBm and monitoring the reflected power. The capability to measure $S_{11}$ at such low power output is critical to prevent saturation of the spin system. Fig. S3 displays the results for the various temperatures we analysed, both without and with green light excitation of the spins (when the static field is on resonance with spins |0⟩↔|1⟩ transition).

The same setup was also used to evaluate the level of spin polarization under light irradiation. This was carried out by comparing the ESR echo signal of the thermal state (without light irradiation) to that of the optically-pumped state. The measurements are carried out both for the |0⟩↔|1⟩ transitions, as well as for the |-1⟩↔|0⟩ transition. With this information, we calculated the populations of the |-1⟩, |0⟩ and |+1⟩ states using the procedure outlined in (*2*), and the results are shown in Fig. S4.

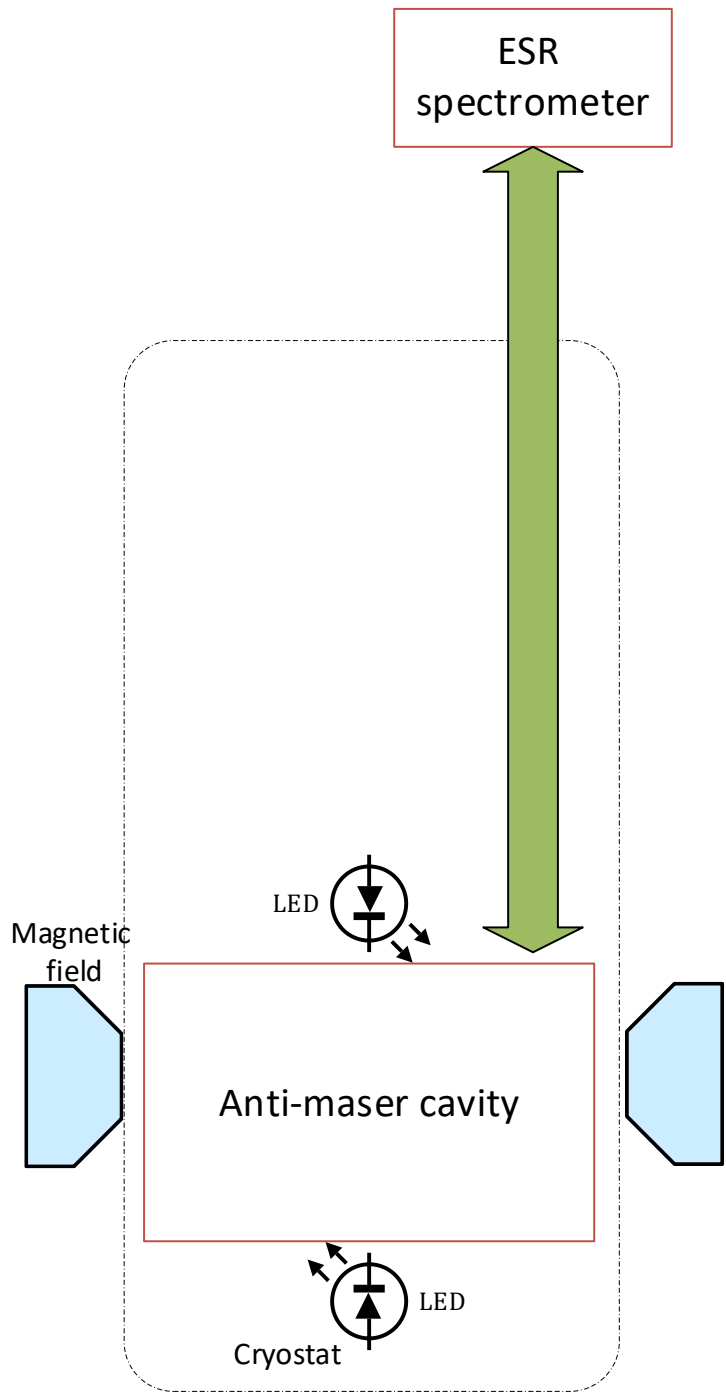

**Figure S2:** Setup for measuring the reflection coefficient of the cavity and also the Zeeman level's population of the NV⁻s at a range of temperatures.

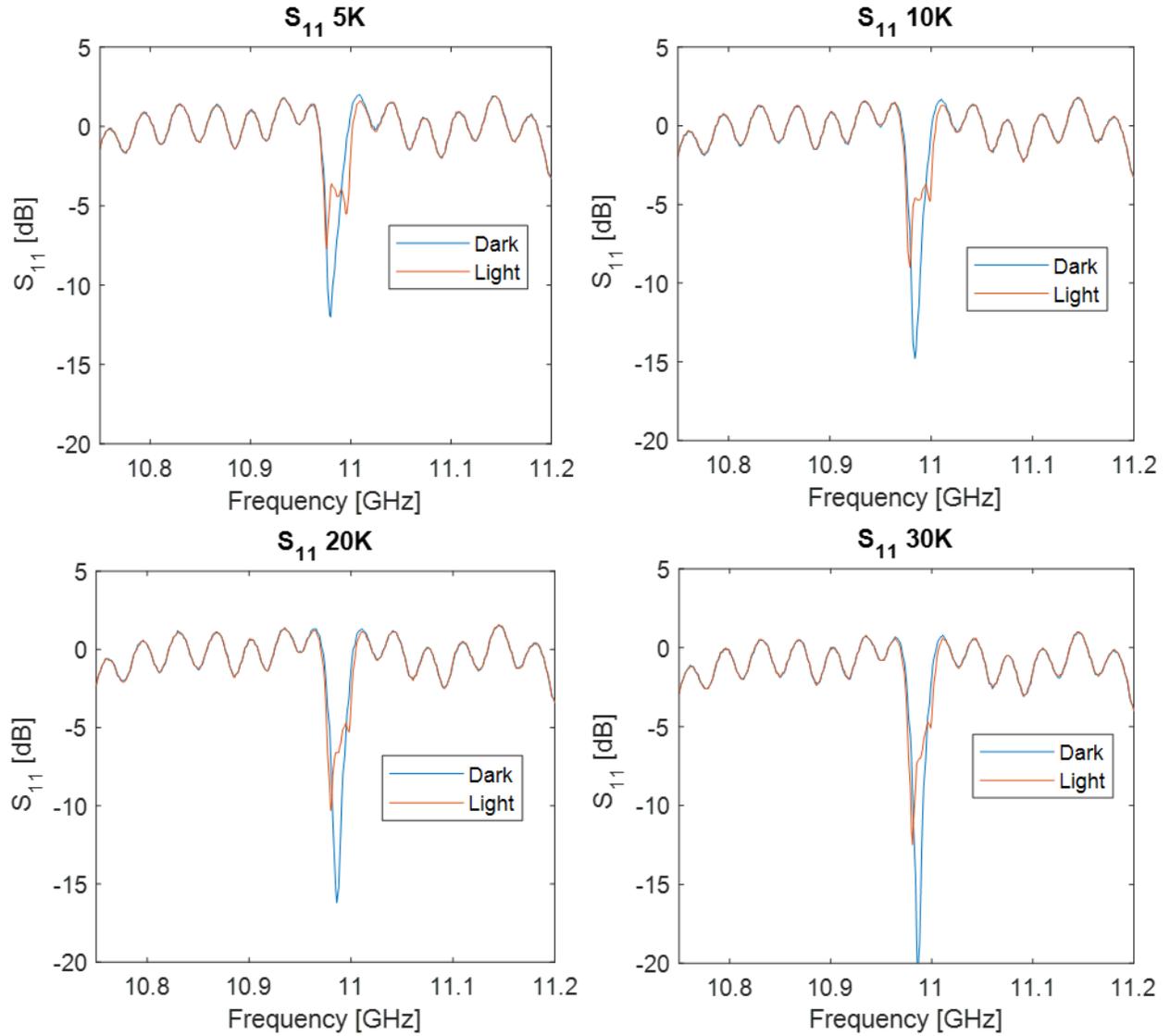

**Figure S3:** Measurement of the reflection coefficient of the anti-maser cavity for different temperatures under light (red) and dark (blue) conditions, when the static magnetic field is tuned to bring the resonance frequency of the spins' $|0\rangle \leftrightarrow |1\rangle$ transition to that of the cavity. When the coupling is close to critical (for example, at 30 K, without light), the coupling rate from the cavity to the outside is $\sim 3.4 \times 10^7$ Hz. At lower temperatures and under light irradiation the resonator is undercoupled, and $S_{11}$ can go up to $\sim$-3.8 dB, which corresponds to coupling rate of $\sim 5.5 \times 10^6$ Hz. It should be noted that due to the limited calibration capability of the reflection coefficient the out-of-resonance wiggles can slightly pass 0 dB, which is within the measurement error range.

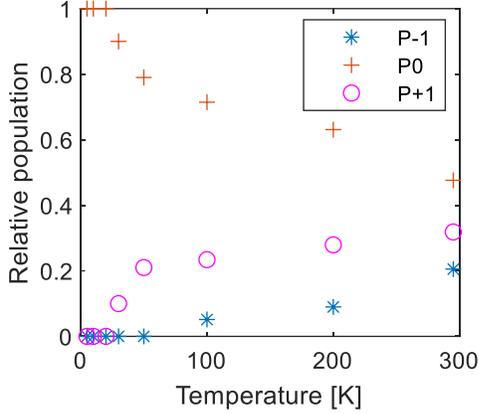

**Figure S4:** The population of the |-1⟩ (blue asterisks), |0⟩ (red pluses) and |1⟩ (magenta circles) Zeeman levels of the NV⁻, under light illumination, as a function of temperature.

## S3. *Measurement of noise power, Y-factor and LNA gain*

*Noise power:* The measurements of the noise power emitted form the cavity, Y-factor and LNA gain were carried out using the experimental setup described in Fig. 2d. The cooled part of the setup is housed in a liquid He-flow cryostat (model STVP-200, Janis, USA). For measuring the noise coming out of the cavity (Figures 3 and S5), point 0 is occupied by a noise source (Keysight model 346CK40) that is turned OFF and thus serves simply as a 50 Ω termination. The noise emitted from the cavity at point 3 goes directly through a directional coupler (with minimal loses of ~ 0.3 dB – see Table S1), enters a low-noise amplifier (Model LNF-LNC6_20A from Low Noise Factory, Sweden), and then goes into the spectrum analyzer (model N9010B, Keysight with built-in pre-amplifier option, bandwidth of 910 kHz). The details of the gains and losses in the setup are provided in Table S1. Examples for the noise power measurements when the 10dB coupler is used in the setup are shown in Fig. S5.

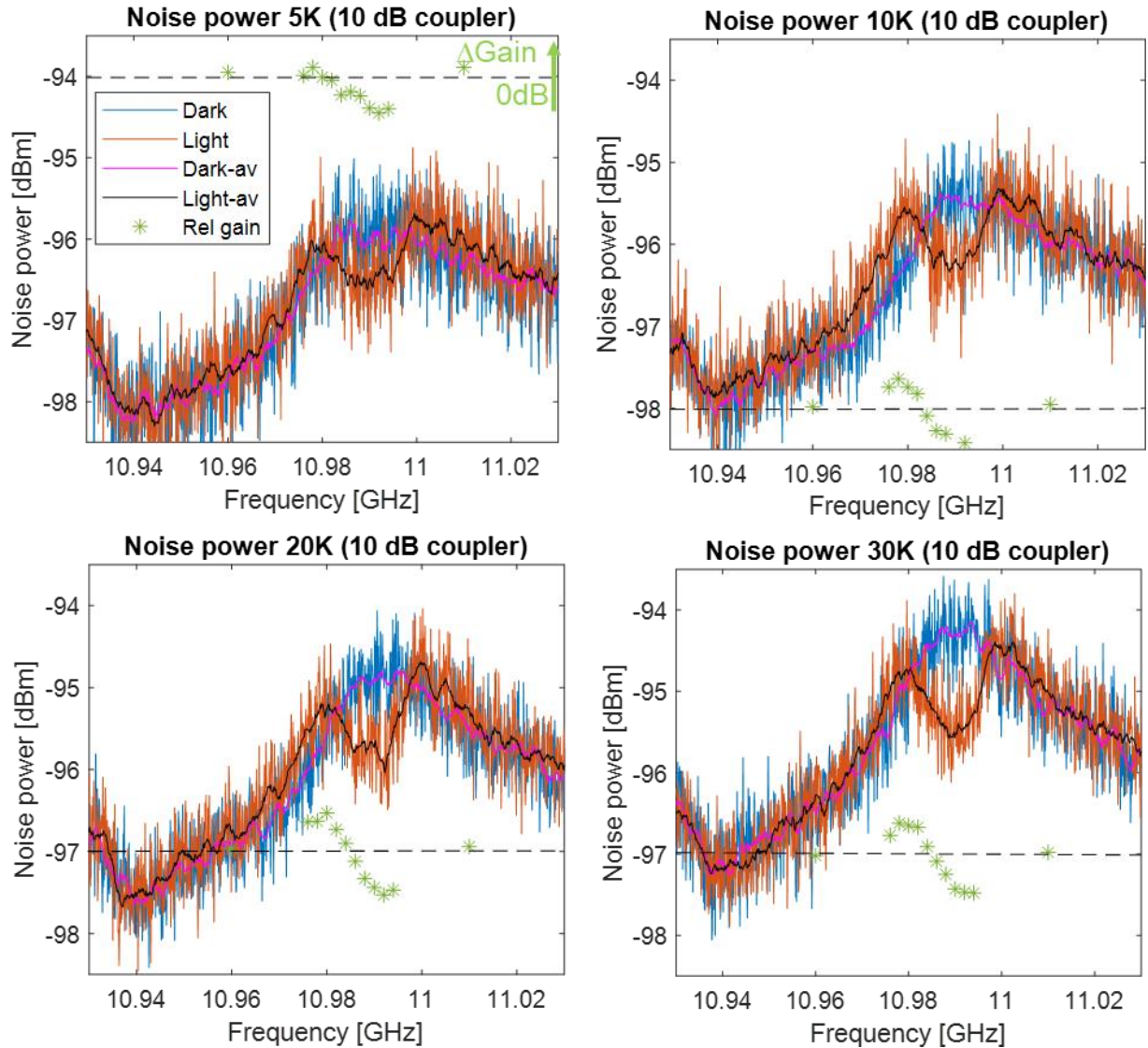

**Figure S5:** Measurement of noise emitted from the anti-maser cavity with a constant addition of noise coming from the 10 dB coupler of our setup (Fig. 2d – noise source is OFF). The data is shown for the temperature range of 5-30 K under dark and light conditions, when the field is set to match the |0⟩↔|1⟩ transition to that of the resonance frequency of the cavity. The raw data noise measurements are superimposed with a moving window average. The green asterisks show the relative change in the gain of the LNA when the light is turned. The scale for the relative gain is in dB according to the left axis but with zero shifted to the dashed line. Legend is the same in all plots.

*Y-factor:* The Y-factor method is a commonly used approach for measuring the noise temperature of microwave devices, particularly amplifiers (*3*). This method involves comparing the noise at a device's output for two distinct input noise levels, typically supplied by a calibrated noise source. This source emits normal thermal noise at ambient temperature when it's off, and substantially higher, calibrated noise levels when on. In our case, the noise source emits 16 dB more noise when on than when off. Ideally, for a noiseless amplifier, the output noise ratio for the ON/OFF states should also be 16 dB. However, in real-world scenarios, this ratio often proves lower due to the inherent noise level of the amplifier or additional noise sources layered onto the original noise from the noise source. By measuring this Y-factor with a Spectrum Analyzer (SA), one can quantify the added noise level at the amplifier input, which in our case, combines the Low Noise Amplifier (LNA) noise and the noise emitted from the cavity. Given that the LNA noise is known from its datasheet, we can use the Y-factor to determine the noise levels emitted from the cavity.

In this work, we enhance the accuracy of our measurements by focusing not on the actual Y-factor value, but on its *change* when light is applied, capitalizing on our knowledge that the noise emitted from the cavity without light correlates with its temperature. For instance, if we refer to Fig. 4 (and Table 1), we observe that at 5 K for a 20 dB coupler, when the light is ON, the Y-factor mostly decreases across the plot but increases by approximately 0.51 dB at the spins' resonance. The minor decrease is attributed to a slight rise in cavity temperature upon light irradiation, while the Y-factor increase results from the reduction of cavity noise at the spins' resonance frequency. We input this value of $\Delta Y$, along with the data from Table S1, and our data on the cavity $S_{11}$ and its variation upon light irradiation, into a MATLAB$^{TM}$ program. This program evaluates various potential cavity noise temperatures upon light irradiation and identifies the one that best corresponds with this change in the Y-factor. The MATLAB$^{TM}$ program is available upon request from the corresponding author and has also been deposited in Zenodo."

*LNA gain:* The gain of the LNA under different input impedance conditions is measured using a calibrated microwave source (SynthHD PRO from Windfreak Technologies, USA) positioned at point 0 in the setup of Fig. 2d. This source transmits a single-frequency tone with a power of -85 dBm to avoid any non-linear effects on the spins. The LNA output is monitored by the Spectrum Analyzer, and changes in gain are primarily observed when the light is toggled ON/OFF, and the static field is in resonance with the spins. No significant changes in LNA gain are observed when the static field is off-resonance and the light is toggled ON/OFF.

| Parameter name | Description | Value and its uncertainty | How is it obtained? |
|---|---|---|---|
| Tn0_OFF | Noise temperature at point 0 when the noise source is OFF | 294 K ± 0.5 K | Noise source data sheet and room temperature |
| Tn0_ON | Noise temperature at point 0 when the noise source is ON | 11760 K ± 20 K | Based on noise source data sheet, Tn1_OFF × 40 |
| $T_{Conn1}$ | Temp. of connector at point 1 and the cables leading to it | 294 K ± 0.5 K | Room temperature |
| $L_{Conn1}$ | Loss of connector at point 1 and the cables leading to it from the noise source | 0.8 dB ± 0.1 dB | Measured at room temperature with VNA |
| Tn1_OFF | Noise temperature at point 1 when the noise source is OFF | 294 K ± 0.5 K | Tn0_OFF × $L_{conn1}$ + (1 - $L_{conn1}$) × $T_{conn1}$ ; ($L_{conn1}$ is in linear scale here) |
| Tn1_ON | Noise temperature at point 1 when the noise source is ON | 9848 K ± 220 K | Tn0_ON × $L_{conn1}$ + (1 - $L_{conn1}$) × $T_{conn1}$ ; ($L_{conn1}$ is in linear scale here) |
| $L_A$ | Power loss of cable A | 0.5 dB ± 0.1 dB | Measured at room temperature with VNA |
| T2 | Physical temperature at point 2 | 5 K ± 1 K | Measured with calibrated AlGaAs diode |
| Tn2_OFF | Noise temperature at point 2 when the noise source is OFF | 277.9 K ± 5.9 K | Numerical integration over the cable's length using Eq. (8-2-4) in (*4*) considering the value of $L_A$ and assuming linear temp gradient along the cable. |
| Tn2_ON | Noise temperature at point 2 when the noise source is ON | 8803.3 K ± 281.3 K | Numerical integration over the cable's length using Eq. (8-2-4) in (*4*) considering the value of $L_A$ and assuming linear temp gradient along the cable. |
| $L_{Coupler}$ | Coupling loss (from point 2 to 5) | 10.1 dB ± 0.1 dB | Measured by VNA at room temperature |
| $T_{Coupler}$ | Physical temperature of the coupler | 5 K ± 1 K | Measured with a calibrated AlGaAs diode |
| T3 | Physical temperature at point 3 | 5 K ± 1 K | Measured with a calibrated AlGaAs diode |
| $T_m$ | Cavity internal noise temperature | 2.5 K ± 0.14 K | The quantity we wish to fit, based on the Y-factor data |
| Γ | Reflection coefficient from the cavity for voltage when light is ON | 0.41 ± 0.019 | $10^{(-S_{11}/20)}$ where $S_{11}$ is measured in real conditions (Fig. S3) |
| Tn3_OFF | Noise coming out of the cavity, | 3.43 ± 0.16 K | $T_m \times (1- \Gamma^2) + T3 \times (\Gamma^2)$ |

| | considering coupling port losses, when noise source is OFF (light is ON) | | **Note that $T_m$ is the quantity we seek in our calculation** |
|---|---|---|---|
| Tn3_ON | Noise coming out of the cavity, considering coupling port losses, when noise source is ON (light is ON) | 3.43 ± 0.16 K | Tn3_OFF |
| $L_B$ | Power loss of cable B | 0.3 dB ± 0.1 dB | Measured at room temperature with VNA |
| $L_{Coupler\_through}$ | Loss of signal going through the coupler from point 4 to 5 | 0.3 dB ± 0.1 dB | Measured by VNA at room temperature |
| Tn5_OFF | Noise temperature at point 5 when the noise source is OFF | 31.06 K ± 0.6 K | [Tn3_OFF × $L_{Coupler\_through}$ × $L_B$ + (1 - $L_{Coupler\_through}$ × $L_B$) × $T_{Coupler}$] × (1-$L_{Coupler}$) + Tn2_OFF × $L_{Coupler}$ ($L_{Coupler\_through}$ and $L_{Coupler}$ are in linear scale here) |
| Tn5_ON | Noise temperature at point 5 when the noise source is ON | 883.6 K ± 28.2 K | [Tn3_ON × $L_{Coupler\_through}$ × $L_B$ + (1 - $L_{Coupler\_through}$ × $L_B$) × $T_{Coupler}$] × (1-$L_{Coupler}$) + Tn2_ON × $L_{Coupler}$ ($L_{Coupler\_through}$ and $L_{Coupler}$ are in linear scale here) |
| G | LNA gain | 31 dB ± 0.5 dB | LNA datasheet, verified with VNA |
| $T_{LNA}$ | Physical temperature of LNA | 5 K ± 1 K | Measured with calibrated AlGaAs diode |
| $TN_{LNA}$ | Noise temperature of LNA | 3 K ± 1 K | Calculated from the LNA datasheet by linear fit where for $T_{LNA}$ = 5 K, $TN_{LNA}$ = 5 K and at $T_{LNA}$ = 295 K, $TN_{LNA}$ = 95 K. |
| Tn6_OFF | Noise temperature at point 6 when the noise source is OFF | 42,887 K ± 1599 K | Tn5_OFF × G (G in linear units) |
| Tn6_ON | Noise temperature at point 6 when the noise source is ON | 1.1162×10$^6$ K ± 25,703 K | Tn5_ON × G (G in linear units) |
| T7 | Physical temperature at point 7 | 294 K ± 0.5 K | Room temperature |
| $L_C$ | Power loss of cable C | 0.5 dB ± 0.1 dB | Measured at room temperature with VNA |
| Tn7_OFF | Noise temperature at point 7 when the noise source is OFF | 38,284 K ± 1,671 K | Numerical integration over the cable's length using Eq. (8-2-4) in (*4*) considering the value of $L_c$ and assuming linear temp gradient along the cable. |
| Tn7_ON | Noise temperature at point 7 when the noise source is ON | $9.9595 \times 10^5$ K $\pm$ 32,372 K | Numerical integration over the cable's length using Eq. (8-2-4) in (*4*) considering the value of $L_c$ and assuming linear temp gradient along the cable. |
| T7 | Physical temperature at point 7 | 294 K $\pm$ 0.5 K | Room temperature |
| $L_{Conn7}$ | Loss of connector at point 7 and the cables leading from it to point 8. | 3.5 dB $\pm$ 0.1 dB | Measured at room temperature with VNA |
| Tn8_OFF | Noise temperature at point 8 when the noise source is OFF | 17,401 K $\pm$ 842 K | Tn7_OFF $\times$ $L_{conn7}$ + (1 - $L_{conn7}$) $\times$ T7 ($L_{conn7}$ is in linear scale here) |
| Tn8_ON | Noise temperature at point 8 when the noise source is ON | $4.486 \times 10^5$ K $\pm$ 17,712 K | Tn7_ON $\times$ $L_{conn7}$ + (1 - $L_{conn7}$) $\times$ T7 ($L_{conn7}$ is in linear scale here) |
| NP | Noise power at SA for 910kHz bandwidth when noise source is OFF | -96.6 dBm $\pm$ 0.2 dB K | $10 \times \log(k_B \times \text{Tn8\_OFF}/0.001 \times 9.1 \times 10^5)$ |

**Table S1:** Example of parameters for evaluating the noise temperature of the cavity due to the anti-maser cooling effect, based on the losses along the microwave path from the noise source to the spectrum analyzer. The data is given for the case where the cavity is at 5K and a directional coupler of 10 dB is used. The numbers of points along the noise path are shown in reference to Fig. 2d.

### S4. *LNA input noise temperature as a function of input impedance*

Low Noise Amplifiers (LNAs) are typically designed to achieve optimal conditions of gain and input noise temperature when the input impedance, as observed from the LNA's perspective, is approximately 50 Ω. It is challenging to measure the dependence of input noise temperature on this parameter, which can assume complex number value, especially at cryogenic temperatures. Therefore, we utilized simulation results graciously provided by the LNA manufacturer, Low Noise Factory. The simulation results, displayed in Fig. S6, predict an optimal noise temperature of roughly 2.5K (for ambient 5K operation), which can degrade up to about 10K when the reflection coefficient (Γ) is approximately -0.8. This can happen, for example, if the amplifier sees an impedance of approximately 5 Ω from its input back, rather than the optimal 50 Ω. In our experiments, the largest reflection coefficient we observed, under light illumination, was |Γ|~ 0.6,

implying that the LNA input noise temperature could degrade up to about 6 K. Due to the large uncertainty as to how much indeed the LNA noise temperature hanged in our experiments, these changes in LNA input noise temperature were not factored into our analysis. Nevertheless, it is evident that the noise temperature of the amplifier does not improve for nonoptimal coupling changes. This suggests that the noise temperatures of the cavity we provide under light illumination are an *upper limit* and in practice, the noise *is lower* than the values we claim as the measured values.

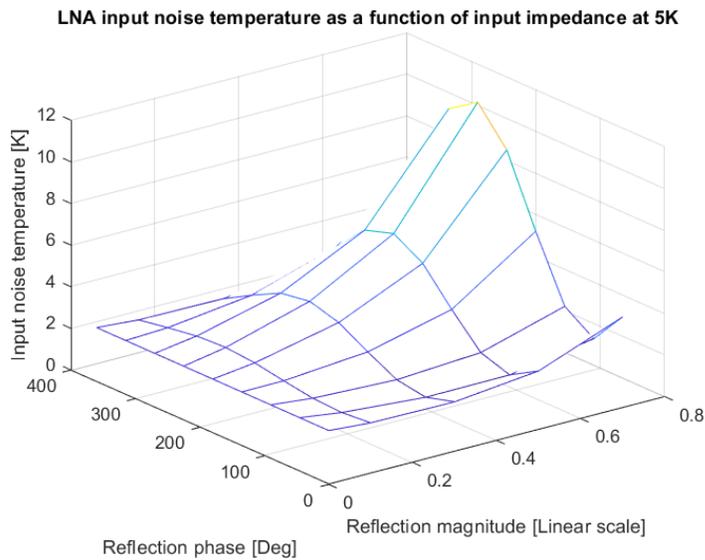

**Figure S6:** Calculated noise temperature for the cryogenic LNA, model LNC6_20 by Low Noise Factory, at 5K, as a function of the reflection coefficient magnitude, as seen from the LNA input back to the line connected to it, for a frequency of 11 GHz.

**References to the SM**

1. A. Sherman *et al.*, Diamond-based microwave quantum amplifier. *Science Advances* **8**, eade6527 (2022).
2. M. Drake, E. Scott, J. Reimer, Influence of magnetic field alignment and defect concentration on nitrogen-vacancy polarization in diamond. *New J Phys* **18**, (2015).
3. Agilent. (2014).
4. A. E. Siegman, *Microwave solid-state masers*. McGraw-Hill electrical and electronic engineering series (McGraw-Hill, New York,, 1964), pp. xv, 583 p.